\documentclass[doublecol, figures]{epl2} 
\usepackage{graphicx}
\usepackage{amsfonts}
\usepackage{amsmath}
\usepackage{epsfig}

\title{Superconducting Fluctuations, Pseudogap and Phase Diagram in
Cuprates}
\shorttitle{Superconducting Fluctuations, Pseudogap and Phase Diagram in
Cuprates}

\author{H. Alloul\inst{1}\thanks{E-mail:\email{alloul@lps.u-psud.fr}} \and F. Rullier-Albenque\inst{2} \and B. Vignolle\inst{3} \and D. Colson\inst{2}\and A. Forget\inst{2}}
\shortauthor{H. Alloul \etal}

\institute{                    
  \inst{1} Laboratoire de Physique des Solides, UMR CNRS 8502, Universit\'e Paris Sud,
91405 Orsay, France\\
  \inst{2} Service de Physique de l'Etat Condens\'e, Orme des Merisiers, CEA Saclay
(CNRS URA 2464), 91191 Gif sur Yvette cedex, France\\
  \inst{3} Laboratoire des Champs Magn\'{e}tiques Intenses, UPR 3228, CNRS-UJF-UPS-INSA, 31400 Toulouse, France}              

\abstract{We report transport measurements using pulsed magnetic fields to suppress
the superconducting fluctuations (SCF) conductivity in a series of YBa$_{2}$Cu$_{3}$O$_{6+x}$ samples.
These experiments allow us altogether to measure
the temperature $T_{c}^{\prime }$ at which SCF disappear, and the pseudogap
temperature $T^{\ast }$. While the latter are consistent with previous
determinations of $T^{\ast}$, we find that $T_{c}^{\prime}$ is slightly
larger than similar data taken by Nernst measurements. A careful
investigation near optimal doping shows that $T^{\ast}$ becomes smaller
than $T_{c}^{\prime}$, which is an unambiguous evidence that the pseudogap
cannot be assigned to preformed pairs. Studies of the incidence of disorder
on both $T_{c}^{\prime}$ and  $T^{\ast}$ allow us to propose a phase
diagram including disorder which explains most observations done in other
cuprate families, and to discuss the available knowledge on the pseudogap
line in the phase diagram.}


\begin{document}

\maketitle

\section{Introduction.}

\begin{figure}
\centering
\includegraphics[width=8cm]{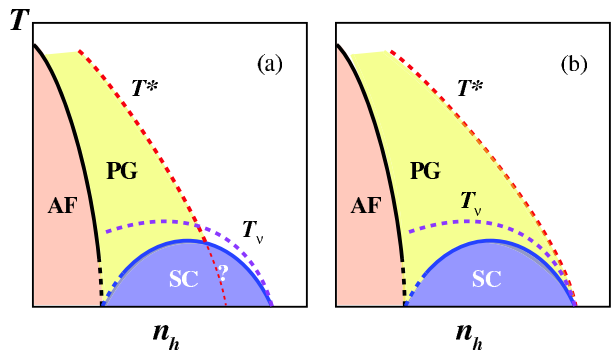}
\caption{(colour on line) Possible phase diagrams for the cuprates. The phase transition
lines delimiting the AF and the SC states are well established. The two
crossover lines lying above $T_{c}$ signal the opening of the pseudogap and
the onset of superconducting fluctuations. Their relative position depends
whether a preformed pair scenario or a competing order applies.}
\label{Fig.1}
\end{figure}

The discovery of High Temperature Superconductivity (HTSC) in the cuprates
has been a turning point in the physics of correlated electron systems.
Superconductivity (SC) happens indeed in a doped Mott insulator for which
nobody would have predicted its occurrence beforehand \cite{PLEE}. While the
search of the "pairing glue" remains a very important aim, it has become
progressively clear during the last 15 years that it would only be possible
to solve this issue after understanding another remarkably robust
experimental property of these systems, the pseudogap, which occurs most
markedly in underdoped systems. The anomalous behaviour of the electronic
properties of these compounds was discovered from a drop of the spin
susceptibility $\chi_{s}(T)$ measured by NMR shifts \cite{Alloul89} below a
temperature $T^{\ast}$ marking the opening of the pseudogap. Specific heat
data \cite{Loram} soon allowed to establish that the density of states
is also strongly reduced. The pseudogap has been found generic, that
is $T^{\ast}$ has a similar variation with doping in bilayer and monolayer \cite{bobroffHg} 
systems. Experimentally it was already clear \cite{Alloul89} that $T^{\ast}$ decreases so abruptly with
increasing doping that it was bound, as shown in Fig.\ref{Fig.1}, either to
intersect the SC dome or to merge with the SC line on the overdoped side of
the phase diagram (PD). Surprisingly, even for such a simple experimental
question, no consensus has been reached so far and the pseudogap line
remains nearly evenly distributed between these two possibilities in most
recent publications.

More importantly, these two types of PD have been associated with radically
different physical explanations of the pseudogap phenomena \cite{PLEE},
either a crossover (or even phase transition) towards a low $T$ correlated
state independent or competing with SC \cite{Emery, Varma} for Fig.\ref{Fig.1}a, 
or a simple precursor pairing state, the pseudogap limiting
then the $T$ range where SC pairs survive as fluctuating entities above 
$T_{c}$, for Fig.\ref{Fig.1}b.

A strong support for a preformed pair scenario has been initially advocated
from the detection of a large Nernst effect \cite{Capan, Ong, RA3} and
of diamagnetism above $T_{c}$ \cite{Wang-PRL2}. Indeed such effects could
be associated with superconducting fluctuations (SCF) and/or vortices
persisting in the normal state. These experiments introduced a new crossover
line which one would like to locate in the phase diagrams of Fig.\ref{Fig.1}. 
As the SCF or vortices require some coherence of the preformed pairs, the
initial proposal that the onset of the Nernst effect occurred at 
$T_{\nu}\approx T^{\ast}/2$, appeared plausible enough to link the pseudogap to
the appearance of preformed pairs. However further data \cite{Wang-PRB2, RA3} 
invalidated this simple relation between $T_{\nu}$ and $T^{\ast}$
and established that $T_{\nu}$ does not follow the increase of $T^{\ast}$
at low doping but rather drops and appears more related to the $T_{c}$ dome,
especially for systems with reduced intrinsic disorder \cite{RA3}.

This inhibits any conclusion concerning the pseudogap to be taken so far
from those experiments. So to discriminate between the PD of Fig.\ref{Fig.1}, 
one rather needs to find external parameters acting differently on $T_{c}$ and the
two crossover temperatures. Many attempts to locate a crossing of the $T^{\ast}$ 
and $T_{c}$ lines have been done by exploiting the robustness of 
$T^{\ast}$ to disorder, established early on \cite{Alloul91, RMP}. The
reduction of $T_{c}$ by Zn substitution allowed through analyses of the
specific heat \cite{Williams} or the NMR shift \cite{Mahajan} to
conclude that, in presence of disorder, any possible $T^{\ast}$ value would
be below $T_{c}$ slightly above optimal doping. This has been taken as
evidence \cite{Williams} that the pseudogap line crosses the SC dome and
reaches $T=0$, which would correspond to a quantum critical point (QCP).
However this conclusion relies on the \textit{assumption} that, even if 
$T^{\ast}$ merges with the onset of superconductivity, impurities would
still suppress $T_{c}$ and not the pseudogap \cite{Mahajan}. But so far no
\textit{reliable experiment has ever established the occurrence of} $T^{\ast}$ 
\textit{values within the SC dome} \cite{RMP}.

The main experimental difficulty remains to distinguish signatures of the
SCF and the pseudogap near optimal doping, both for macroscopic probes
and for spectral studies by superconducting tunneling (STM) \cite{Renner} or Angle Resolved Photoemission (ARPES) \cite{Ding}. 

In the present work we therefore use a novel experimental approach we
established recently \cite{RA-HF}, which allowed us to suppress the
SCF contribution to the ab plane conductivity in single crystal samples
with large pulsed magnetic fields. We do then
delineate the onset $T_{c}^{\prime}$ of the SCF, which we find located
slightly above $T_{\nu}$ and recover the $T$ variation of the normal state
resistivity down to $T_{c}$ from which $T^{\ast}$ is determined
independently. We exploit here this possibility to determine both 
$T_{c}^{\prime}$ and $T^{\ast}$ crossover lines with a single experiment 
and do give evidence that they intersect around optimal doping, which
prohibits considering the pseudogap as the universal onset of pair formation, 
and favours the PD of Fig.\ref{Fig.1}a. Our results allow us then to discuss the experimental
significance of the pseudogaps detected by the spectroscopic experiments. 

\section{Samples and techniques}

Single crystals of YBa$_{2}$Cu$_{3}$O$_{6+x}$ (YBCO) with various oxygen contents 
have been grown and electrically contacted as described elsewhere \cite{RA-EPL1}. 
We have studied two underdoped samples (labelled as UD57 and UD85) with $T_{c}$ 
(taken at the midpoint of the resistive transition) of 57.1 and 84.6K corresponding 
approximatively to oxygen contents of 6.54 and 6.8 respectively, an optimally-doped 
sample (OPT93.6) with $T_{c}=93.6$K and a slightly overdoped sample (OD92.7) with $T_{c}=92.7$K. 
The hole doping $n_{h}$ in these different samples is estimated from the parabolic 
relationship between $T_{c}/T_{c}^{max}$ and $n_{h}$ \cite{Tallon-Tcp}.

\begin{figure}
\centering
\includegraphics[width=8cm]{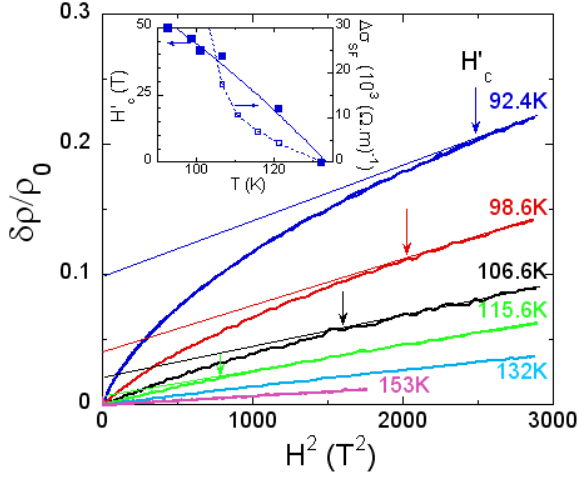}
\caption {(colour on line) Field variation of $\delta\rho/\rho_{0}$ plotted versus $H^{2}$
for decreasing temperatures down to $T\approx T_{c}$ in the UD85 sample. For $T>130$K 
the magnetoresistance follows an $H^{2}$ dependence, but at lower $T$
the contribution of superconducting fluctuations increases the low field
conductivity. This occurs below an onset field $H_{c}^{\prime}(T)$ (arrow)
which is plotted in the inset together with $\Delta\sigma_{SF}(T,0)$.
Both quantities become negligible at a temperature $T_{c}^{\prime }=130$K,
which defines the onset of SCF.}
\label{Fig.2}
\end{figure}

Let us first recall the method we have proposed \cite{RA-HF} to
recover in high fields the magnetoresistance (MR)associated with the normal
state quasiparticles. In a single band metal the transverse orbital MR can be written as 
\begin{equation}
\delta \rho /\rho _{0}=\frac{\rho (T,H)-\rho (T,0)}{\rho (T,0)}=
(\omega _{c}\tau )^{2}  \label{Eq.1}
\end{equation}
where $\omega_{c}=eH/m^{\ast}$ is the cyclotron frequency and $\tau$ an
electronic scattering time, so that, as shown by Harris et al \cite{Harris},
the $H^{2}$ dependence expected in the low field limit $\omega _{c}\tau
<<1$ is obtained in the cuprates at high $T$ for moderate low fields. We
have extended this approach by performing $\rho(T,H)$ measurements, using
pulse fields as large as 55 Tesla. In view of the required accuracy, we
have eliminated any Hall contribution to the measured voltage by reversing
the magnetic field. The $H^{2}$ dependence could be seen to persist in
fields as large as 55 Tesla in the cuprates as long as $T>>T_{c}$. As
exemplified in Fig.\ref{Fig.2} for the UD85 sample, $\delta\rho/\rho _{0}$ 
indeed displays an $H^{2}$ variation for $T>130$K, which
indicates that the weak field limit still applies up to 55T.

However large departures with respect to this quadratic behaviour appear
when $T$ is lowered towards $T_{c}$. As already stated in the case of the UD57 
\cite{RA-HF}, we associate the initial fast increase of $\delta\rho/\rho _{0}$ with the
applied field to the destruction by the applied field of the
paraconductivity contribution. In such a case the normal state magnetoresistance
is only recovered at fields exceeding a $T$ dependent threshold field 
$H_{c}^{^{\prime}}$. This experimental approach allows us then to single out
the normal state properties and separate the contributions of SCF to the in
plane transport.

\section{Determination of $T_{c}^{\prime}$}

To analyse then the data, it appears quite natural to apply a two fluid
model, for which, at a given $T$, the conductivities due to the normal state
quasi-particles $\sigma _{n}(H)$ and that of the superconducting
fluctuations $\Delta\sigma _{SF}(H)$ are additive, that is
\begin{equation}
\rho^{-1}(H)=\sigma (H)=\sigma _{n}(H)+\Delta\sigma_{SF}(H)  
\label{Eq.2}
\end{equation}
At a given $T$, the limiting high field behaviour $\delta\rho/\rho_{0}%
=\delta\rho_{n}/\rho_{0}+BH^{2}$ displayed in Fig.\ref{Fig.2} allows us
then to determine
\begin{equation}
\rho_{n}(H)=\rho_{0}+\delta\rho_{n}+B\rho_{0}H^{2}=\sigma_{n}^{-1}(H).
\label{Eq.3}
\end{equation}
Extending such an analysis to the various $T$ values for which the normal
state can be recovered with 55Tesla applied field allows us then to
determine both $\rho_{n}(T,H)$ and $\Delta\sigma_{SF}(T,H)$ from Eq.\ref{Eq.2} 
and Eq.\ref{Eq.3}. Here we shall concentrate on the conditions for
which $\Delta\sigma_{SF}(T,H)$ is fully suppressed, that is for $H>H_{c}^{\prime}(T)$. 
It can be seen in Fig.\ref{Fig.2} that $\Delta\sigma_{SF}(T,0)$ vanishes 
for increasing $T$ at a temperature $T_{c}^{\prime}$
which naturally coincides with that for which $H_{c}^{\prime}(T)$ vanishes. 

\begin{figure}
\centering
\includegraphics[width=8cm]{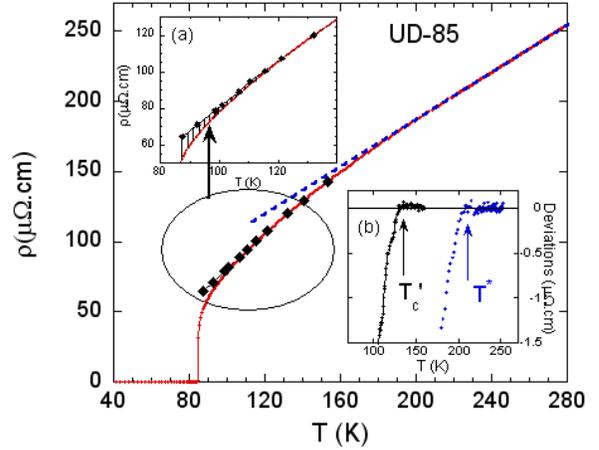}
\caption{(colour on line) $T$ variation of the resistivity of the UD85 
sample studied. The full square data for $\rho_{n}(T,0)$
are deduced here from data of Fig.2.
The enlargement in inset(a) better exhibits the
SCF contribution (hatched area). In inset(b) the
resistivity decrease with respect to the high $T$ linear extrapolation 
(dashed line in the main panel) due to the
opening of the pseudogap is displayed together with that due to SCF.}
\label{Fig.3}
\end{figure}

This procedure also allows us to extrapolate the zero field resistivity 
$\rho_{n}(T,0)=\rho_{0}+\delta\rho_{n}$ in the absence of SC
fluctuations, which is identical to $\rho(T,0)$ down to $T_{c}^{\prime}$.
As an example, the analysis of the UD85 sample data of Fig.\ref{Fig.2} 
yields the $\rho_{n}(T,0)$ given in Fig.\ref{Fig.3}. 

Let us note that in the pure UD57 sample $T_{c}^{\prime}$ has been
found larger \cite{RA-HF} than the onset of $T_{\nu}$ found for
the same samples \cite{RA3}, which reflects the fact that the SCF regime is
a crossover towards normal state behaviour. The present method is then  a
quite sensitive experimental approach to study the magnitude of the SCF. The raw
data similar to those in Fig.2 and the
detailed analysis of the $T$ and $H$ dependences of $\Delta\sigma _{SF}$
will be discussed in a full report \cite{FRATBP}.

\section{Normal state resistivity versus $T$, determination of $T^{\ast}$}

In transport measurements, it has often been assumed that the pseudogap can
be determined from the downward departure of $\rho_{ab}$ from its linear
high $T$ variation \cite{Ito, Wuyts, Raffy}. 
In very underdoped cases such as those of UD85 or UD57
one can determine directly $T^{\ast}$ without considering
the occurrence of SCF at $T_{c}^{\prime}$, as those two
crossovers are quite well separated. For instance in UD85, 
the resistivity deviates from linearity below $T^{\ast}\simeq 210$K, 
well above $T_{c}^{\prime}\simeq $130K as
evidenced in Fig.\ref{Fig.3}. This value of $T^{\ast}$ agrees
well with that ($\simeq $250K) obtained from $^{89}$Y NMR shift data for similar
oxygen contents \cite{Alloul91, Tallon-loram}. 

\begin{figure}
\centering
\includegraphics[width=8cm]{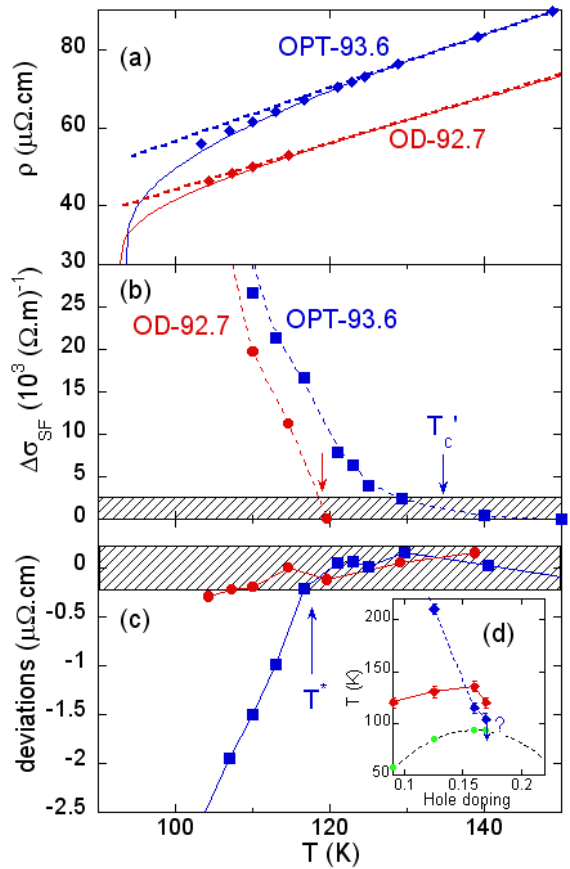}
\caption{(colour on line)(a) $T$ variations of the zero field resistivities and $\rho_{n}(T,0)$ 
deduced from high field data for OPT-93.6 and
OD92.7 samples.(b)Onset of the SCF contribution to the
conductivity allowing to determine $T_{c}^{\prime}$. (c) Deviations from
linearity of $\rho_{n}(T,0)$ allowing to determine the pseudogap $T^{\ast}$. 
(d) Phase diagram versus hole doping showing
that the crossover lines for $T_{c}^{\prime}$ and $T^{\ast}$ intersect
around optimal doping.}
\label{Fig.4}
\end{figure}

However, more care is required for larger dopings, when $T^{\ast}$ approaches $T_{c}$. 
If one considers the zero field
transport data given in Fig.\ref{Fig.4}a for OPT93.6 and OD92.7, 
the slopes of the linear $T$ dependences of $\rho(T)$ are quite
well defined, and distinct downward deviations are apparent. Those could be associated 
with the pseudogap  $T^{\ast}$ if one were to ignore the SCF. 
However, as can be seen in Fig.\ref{Fig.4}b,
the data for $H_{c}^{\prime}(T)$ or $\Delta\sigma _{SF}(T,0)$ give evidence 
that SCF are the primary source for these
downturns, with  $T_{c}^{\prime}$ values of 135(5) K and 120(5)K
respectively (in agreement with diamagnetic measurements for the OPT sample \cite{Li}). 
But one can see on Fig.\ref{Fig.4}a that, when the
contribution of SCF is suppressed, the data for $\rho_{n}(T,0)$ still
displays a downward curvature due to the pseudogap. 
This is better displayed in Fig.\ref{Fig.4}c, in which the deviation of $\rho_{n}(T,0)$ from linearity
has been magnified. There one can determine $T^{\ast}=118(5)$K for OPT-93.6, 
while one cannot ascertain any deviation for OD92.7
within experimental accuracy, so that $T^{\ast}<100$K in that case. 
Notice that we consistently defined here $T_{c}^{\prime}$ 
and $T^{\ast}$ at temperatures for which deviations of $\Delta\sigma _{SF}(T,0)$ or 
$\rho_{n}(T,0)$ correspond to the same magnitude (hatched areas in Fig.\ref{Fig.4}b and c).

So, while $T_{c}^{\prime}$ is always smaller than $T^{\ast}$ for
underdoped samples, the present data give evidence that $T^{\ast%
}<T_{c}^{\prime}$ for OPT93.6, while for OD92.7 any
possible value for $T^{\ast}$ would only also occur below $T_{c}^{\prime}$. 
This establishes that the pseudogap line crosses the onset of SCF for a
hole content slightly below optimal doping, as shown in Fig.\ref{Fig.4}d, 
where the data for $T_{c}^{\prime}$ and $T^{\ast}$ are
summarized versus the equivalent hole contents. This result definitively \textit{prohibits considering that the pseudogap could be the onset of pairing}. We shall discuss its importance after showing the incidence of disorder on these two lines in the phase diagram.

\section{Influence of disorder}

It is well established that disorder strongly depresses $T_{c}$ 
while $T^{\ast}$ is not affected \cite{Alloul91, RMP}. This can be directly evidenced from resistivity measurements in electron irradiated crystals. Electron irradiation at low $T$ has been shown to introduce homogeneously distributed point defects without changing the hole doping \cite{RA1, RA2}. 

\begin{figure}
\centering
\includegraphics[width=8cm]{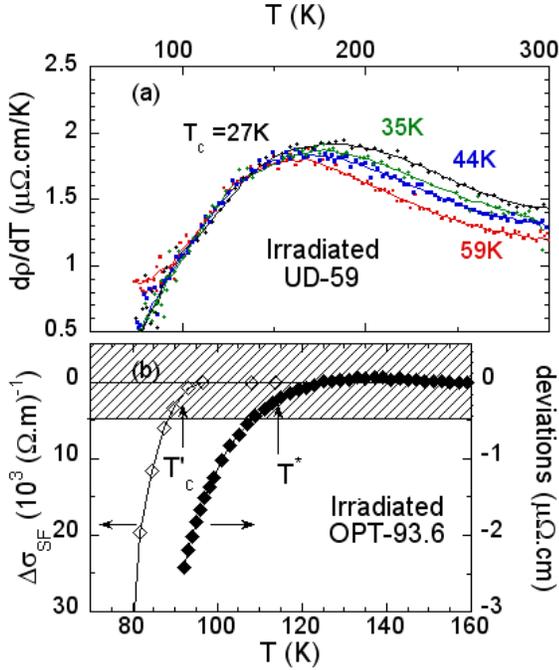}
\caption{(colour on line)(a) Variations of $d\rho_{ab}/dT$ versus temperature in the
pure UD59 sample and after electron irradiation at low $T$. The maximum 
of $d\rho_{ab}/dT$, which roughly 
corresponds to $T^{\ast}/2$ does not vary markedly with disorder. (b)
Similarly in OPT93.6, pulse field measurements gives evidence that 
$T_{c}^{\prime}$ is depressed to 92K by disorder, while $T^{\ast}$ 
remains nearly unmodified at $\approx 114$K. We have used the same criterion 
as in fig.4b and c to define $T_{c}^{\prime}$ and $T^{\ast}$ (hatched area) although better data accuracy is available here as $T^{\ast}$ is obtained directly from zero field measurements ).}
\label{Fig.5}
\end{figure}

For an UD59 single crystal, Fig.\ref{Fig.5}a shows that $d\rho_{ab}/dT$ increases 
with decreasing $T$ below 300K, which would correspond to $T^{\ast}>300$K, in agreement  with $T^{\ast%
}\simeq 350$ to $400$K found from NMR \cite{Alloul89, Alloul91, Williams} for all
samples with oxygen contents between 6.6 and 6.7 at the
$T_{c}\approx$ 60K plateau. In such samples one cannot use the deviation of $\rho_{ab}(T)$ 
from linearity to determine $T^{\ast}$, as diffusion of oxygen takes place above room
temperature yielding significant variations of the chain contributions to 
$\rho(T)$ \cite{Lavrov}. However it has  been shown that the inflexion point 
in $\rho(T)$ is located at $T_{1}\simeq T^{\ast}/2$ \cite{Wuyts, Raffy}\footnote{Ando et al. \cite{Ando} propose another determination of $T^{\ast}$ based on the variation of the second derivative of the resistivity. This leads to much lower values of $T^{\ast}$ than those determined by NMR.}. 
The fact that this point only slightly increases 
with increasing defect content is a good confirmation of the NMR results that $T^{\ast}$ is not modified by disorder . This is also a different phrasing of the evidence that Matthiessen's rule applies in the normal state of cuprates \cite{RA1}. 
On the contrary, we have previously shown from Nernst effect data \cite{RA3} and 
high field measurements \cite{RA-HF}
that disorder depresses $T_{\nu}$ and $T_{c}^{\prime}$, although in much less proportion than $T_{c}$.

We have also performed pulse field measurements on an
OPT-93.6 sample with $T_{c}$ reduced down to 70.7K by electron irradiation in order 
to test the influence of disorder on both $T_{c}^{\prime}$ and $T^{\ast}$ when these
temperatures become comparable. 
As can be seen in  Fig.\ref{Fig.5}b, $T_{c}^{\prime}$ is found there to decrease 
markedly down to 92(4)K. Therefore, above 92K, the magnetoresistance remains quadratic in magnetic field and the data for $\rho(T)$ is not influenced
by SCF for this sample. The detected deviation from the high $T$ linear
behaviour gives $T^{\ast}$=114(6)K, in good agreement with
that obtained above in the pure sample after suppressing the SCF in high
fields. So we confirm here that defects
depress $T_{c}$ and $T_{c}^{\prime}$, but
have a much smaller incidence on $T^{\ast}$, which shows that these two crossover
temperatures are associated with independent physical phenomena.

\section{Discussion}

The use of high applied fields has been essential in allowing us to depress
the SCF in pure YBCO samples, which permits us to shed
some light on the continuing debate concerning the pseudogap and SCF in
cuprates.  We do demonstrate that over the hole concentration range from
underdoped to slightly overdoped, the SCF crossover $T_{c}^{\prime}$ and $\Delta\sigma_{SF}$ are quite similar. Furthermore $T_{c}^{\prime}$ is affected in the same manner by large
applied fields. On the contrary the pseudogap crossover $T^{\ast}$ is shown
to depend markedly on doping and to cross $T_{c}^{\prime}$ at optimal
doping, so that the onset of pairing occurs there above $T^{\ast}$. Here we
have consistently located $T^{\ast}$ and $T_{c}^{\prime}$ for different
dopings at temperatures corresponding to resistivity deviations of the same
magnitude.

\begin{figure}
\centering
\includegraphics[width=8cm]{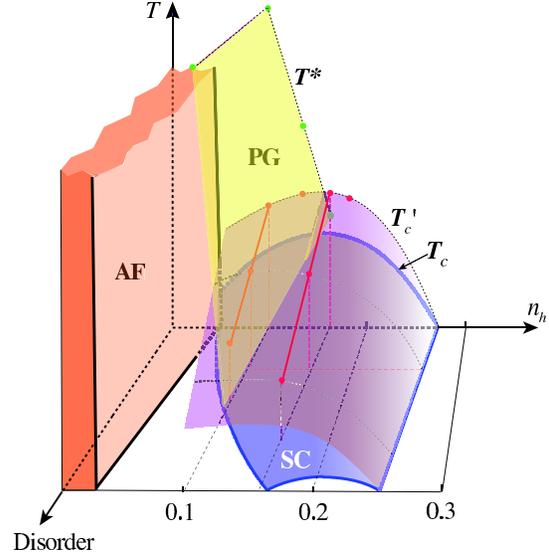}
\caption{(colour on line) Phase diagram constructed on the data points
obtained here, showing the evolution of $T_{c}^{\prime}$ the onset of
SCF, with doping and disorder. The pseudogap and SCF
surfaces intersect each other near optimum doping in the clean limit. 
These surfaces have been limited to ranges where
they have been determined experimentally. In the overdoped regime, data taken on Tl 2201 indicates
that disorder suppresses SC without any anomalous extension of the SCF \cite{RA1}.}
\label{Fig.6}
\end{figure}

Furthermore we confirmed, within the same experimental procedure, that
disorder does not affect $T^{\ast}$ while it depresses $T_{c}^{\prime}$,
so that the clean case phase diagram displayed in the inset of Fig.\ref%
{Fig.4}c can be extended in a third dimension by introducing a disorder axis
in Fig.\ref{Fig.6}, for realistic ranges of disorder. There we display that
i) the $T_{c}$ depression is faster for underdoped samples than for optimal
ones, ii) the $T^{\ast}$ line does not evolve with disorder, iii) the onset 
$T_{c}^{\prime}$ of SCF decreases at a much slower rate in the underdoped
case than for optimal doping. 

Consequently the SCF regime appears then much more extended with respect to $%
T_{c}$ values in the underdoped samples when a large disorder is introduced.
Such fluctuations were even detected in underdoped samples for which $T_{c}$
was nearly reduced to zero \cite{RA-HF}. This PD therefore mimics the situation
encountered for Nernst data taken in La$_{2-x}$Sr$_{x}$CuO$_{4}$, for which
the SCF regime was found to be peaked in the underdoped range \cite{Ong}. This
agrees with our former proposal that large intrinsic disorder explains the metal to insulator crossover seen in that case. For similar dopings, YBCO remains in a well defined metallic state at low $T$ \cite{RA-EPL2}.

The specific behaviours of the SCF and of the pseudogap remain among the
most important features of the cuprate physics. It is quite clear that,
independently of the actual possible origin of the pseudogap, SCF are
present above the 3D $T_{c}$ in these 2D systems with short coherence
length. A detailed coherent analysis of these fluctuations will be performed elsewhere
\cite{FRATBP}.

As for the pseudogap phenomenon, anomalies of distinct transport experiments
have been detected above $T_{c}$ and assigned as well to the pseudogap, 
locating it at much higher $T^{\ast}$ values for optimal doping than found
here. This has been the case for an increase in the c axis conductivity \cite%
{Shibauchi} or for the onset of in plane anisotropy of the Nernst
coefficient \cite{Daou} in YBCO samples. While in plane transport and
magnetic susceptibilities have been studied at length in a variety of
cuprates for the last twenty years, these recent experiments are less understood
and might as well be unable to separate the SCF from the pseudogap. The reproducibility of the detected anomalies in other HTSC, and
their insensitivity to disorder have not been established so far, so that
their use to locate $T^{\ast}$ does not appear to us as robust as the
present approach. 

Meanwhile STM \cite{Renner} or  ARPES \cite{Ding} spectroscopic experiments, 
have detected gaps in the energy spectrum, which are found to persist above $T_{c}$
in underdoped but also in overdoped cuprates .
The excited states detected by STM experiments are easier to
analyze in the overdoped cuprates in the absence of pseudogap \cite{Kugler, Yazdani}. 
The d-wave superconducting gap detected at low $T$ does not fully
disappear above the 3D $T_{c}$, so that the remaining dips in the energy
spectrum can only be associated with SCF in this doping
range. 

For optimally doped samples the SCF may extend below or above the
onset of the pseudogap, as shown here. But it is not possible so far to use
large enough fields to separate them in the spectroscopic ARPES and STM
experiments. This explains why remnants of the SC gap due to SCF could be
often taken inadvertently as "pseudogaps".

We have shown that the robustness of the pseudogap to disorder is an
important clue to separate its spectral features from those associated with
SC. In the STM and ARPES experiments for which surface inhomogeneities are
apparently extremely important \cite{Lang-Davis}, it seems to us that the large
gap detected in underdoped samples ought to be associated with the
pseudogap, in view of its robustness and reproducibility. The Fermi arcs (or
pockets) are also undoubtedly experimental features associated with the
pseudogap state.

The pseudogap being distinct from SCF, one then expects to detect different spectral 
responses by STM or ARPES for these two phenomena below optimal
doping. But for underdoped samples experimental difficulties could
have been anticipated from the early experiments on the cuprates. Indeed,
NMR experiments have shown for long that the depression of $\chi_{s}$ at
low $T$ is so large that it was nearly impossible to detect the further
reduction of $\chi_{s}$ expected below $T_{c}$. So if the large gap
detected in the underdoped systems is indeed associated with the pseudogap,
it renders quite difficult the detection of spectral modifications occurring
then below $T_{c}^{\prime}$. This explains the debate between one gap
\cite{Kanigel} and two gap scenarios \cite{Le Tacon, Lee-Shen} which has been
engaged for a while, and for which no consensus has emerged so far.

Finally, if any feature detected above $T_{c}$ is not always connected with
the pseudogap, one may wonder as well if anything detected below $T^{\ast}$
is an essential phenomenon or a secondary transition which might be non
universal to the cuprate physics and driven by specific system details. If a competing
order which breaks time reversal symmetry occurs somewhat below the $%
T^{\ast}$ determined here \cite{Kaminski, Fauqué, Xia}, one still needs to establish the
correlation between the various experimental aspects of the pseudogap -loss
of spin susceptibility, transfer of spectral weight from low to high
energies, Fermi arc observations,...- and the symmetry breaking.

\begin{acknowledgments}
This work was supported by ANR grant NT05-4\_41913. We
thank C. Proust for support during experiments at LNCMP-Toulouse which have been funded 
by the FP6 "Structuring the European Research
Area, Research Infrastructure Action" contract R113-CT2004-506239. 
\end{acknowledgments}


\begin{thebibliography}{99}
\bibitem{PLEE}
\Name{Lee P. A.,\ Nagaosa N. \and Wen X.G.}
\Review{RMP} {78} {2006} {17}.

\bibitem{Alloul89} 
\Name{Alloul H.,\ Ohno T. \and Mendels P.}
\Review{Phys. Rev. Lett.} {63} {1989} {1700}.

\bibitem{Loram}
\Name{Loram J. W.} {et al.}
\Review{Phys. Rev. Lett.} {71} {1993} {1740}. 

\bibitem{bobroffHg}
\Name{Bobroff J.}  {et al.}
\Review{Phys. Rev. Lett.} {78} {1997} {3757}.

\bibitem{Emery}
\Name{Emery V.J. \and Kivelson S.A.}
\Review{Nature} {374} {1995} {434}.

\bibitem{Varma}
\Name{Varma C.M.}
\Review{Phys. Rev. B} {55} {1997} {14554}; 
\Review{Phys. Rev.Lett.} {83} {1999} {3538}.

\bibitem{Capan} 
\Name{Capan C.} {et al.}
\Review{Phys. Rev. Lett.} {88} {2002} {056601}.

\bibitem{Ong}
\Name{Wang Y.} {et al.}
\Review{Science} {299} {2003} {86}.

\bibitem{RA3} 
\Name{Rullier-Albenque F.} {et al.}
\Review{Phys. Rev. Lett.} {96} {2006} {067002}.

\bibitem{Wang-PRL2}
\Name{Wang  Y.} {et al.}
\Review{Phys. Rev. Lett.} {95} {2005} {247002}.

\bibitem{Wang-PRB2} 
\Name{Wang Y.,\ Li L. \and Ong N.P.}
\Review{Phys. Rev. B} {73} {2006} {024510}.

\bibitem{Alloul91}
\Name{Alloul H.} {et al.}
\Review{Phys. Rev. Lett.} {67} {1991} {3140}.
 
\bibitem{RMP}
\Name{Alloul H.} {et al.}
\Review{Rev. Mod. Phys.} {81} {2009} {45}.

\bibitem{Williams}
\Name{Williams G.V.M.} {et al.}
\Review{Phys.Rev. Lett.} {78} {1997} {721}.

\bibitem{Mahajan}
\Name{Mahajan A.V.} {et al.}
\Review{Phys. Rev. Lett.} {72} {1994} {3100}.

\bibitem{Renner}
\Name{Renner Ch.} {et al.}
\Review{Phys. Rev. Lett.} {80} {1998} {149}.

\bibitem{Ding}
\Name{Ding  H.} {et al.}
\Review{Nature} {382} {1996} {51};
\Name{Marshall D. S.}  {et al.}
\Review{Phys. Rev. Lett.} {76} {1996} {4841}.

\bibitem{RA-HF}
\Name{Rullier-Albenque F.} {et al.}
\Review{Phys. Rev. Lett.} {99} {2007} {027003}.

\bibitem{RA-EPL1}
\Name{Rullier-Albenque F.} {et al.}
\Review{Europhys. Lett.} {50} {2000} {81}. 

\bibitem{Tallon-Tcp}
\Name{Tallon J.L.} {et al.}
\Review{Phys. Rev. B} {51} {1995} {12911}.

\bibitem{Harris}
\Name{Harris J.M.} {et al.}
\Review{Phys. Rev. Lett.} {75} {1995} {1391}.

\bibitem{FRATBP}
\Name{Rullier-Albenque F.} {et al.}
in preparation.

\bibitem{Ito}
\Name{Ito T.} {et al.}
\Review{Phys. Rev. Lett.} {70} {1993} {3995}.

\bibitem{Wuyts}
\Name{Wuyts B.,\ Moshchalkov V. V. \and Bruynseraede Y.}
\Review{Phys. RevB} {53} {1996} {9418}.

\bibitem{Raffy}
\Name{Konstantinovic Z.,\ Li Z. \and Raffy H.}
\Review{Physica C} {351} {2001} {1297}.

\bibitem{Tallon-loram}
\Name{Tallon J.L. \and Loram J.W.}
\Review{Physica C} {349} {2001} {53}.

\bibitem{Li}
\Name{Li L.} {et al.}
\Review{Phys. Rev.B} {81} {2010} {054510}.

\bibitem{RA1}
\Name{Rullier-Albenque F.,\ Alloul H. \and Tourbot R.}
\Review{Phys. Rev. Lett.} {87} {2001} {157001}.

\bibitem{RA2}
\Name{Rullier-Albenque F.,\ Alloul H. \and Tourbot R.}
\Review{Phys. Rev. Lett.} {91} {2003} {047001}.

\bibitem{Lavrov}
\Name{Lavrov A.N. \and Kozeeva L.P.}
\Review{Physica C} {253} {1995} {313} and references therein.

\bibitem{Ando}
\Name{Ando Y.} {et al.}
\Review{Phys. Rev. Lett.} {93} {2004} {267001}.

\bibitem{RA-EPL2}
\Name{Rullier-Albenque F.} {et al.}
\Review{Europhys. Lett.} {81} {2008} {37008}.

\bibitem{Shibauchi}
\Name{Shibauchi T.} {et al.}
\Review{Phys. Rev. Lett.} {86} {2001} {5763}.

\bibitem{Daou}
\Name{Daou R.} {et al.}
\Review{Nature} {463} {2010} {519}. 

\bibitem{Kugler}
\Name{Kugler M.} {et al.}
\Review{Phys. Rev. Lett.}{86} {2001} {4911}.

\bibitem{Yazdani}
\Name{Gomes K.K.} {et al.}
\Review{Nature} {447} {2007} {569}.

\bibitem{Lang-Davis}
\Name{Lang K.M.} {et al.}
\Review{Nature} {415} {2002} {412}.

\bibitem{Kanigel}
\Name{Kanigel A.} {et al.}
\Review{Phys. Rev. Lett.} {101} {2008} {137002}.

\bibitem{LeTacon}
\Name{Le Tacon M.} {et al.}
\Review{Nature Physics} {2} {2006} {537}.

\bibitem{Lee-Shen}
\Name{Lee W.S.} {et al.}
\Review{Nature} {450} {2007} {81}.

\bibitem{Kaminski}
\Name{Kaminski A.} {et al.}
\Review{Nature} {416} {2002} {610}.

\bibitem{Fauqué}
\Name{Fauqu\'e B.} {et al.}
\Review{Phys. Rev. Lett.} {96} {2006} {197001}.

\bibitem{Xia}
\Name{Xia J.} {et al.}
\Review{Phys. Rev. Lett.} {100} {2008} {127002}. 

\end{thebibliography}
\end{document}